\documentclass[conference]{IEEEtran}
\IEEEoverridecommandlockouts
\input{include/customize.tex}
\input{./include/gloss.tex}

\usepackage{amsmath,amssymb,amsfonts}
\usepackage{algorithmic}
\usepackage{graphicx}
\usepackage{textcomp}

\newcommand{\chapter}{\section}
\renewcommand{\vec}[1]{\boldsymbol{#1}}

\begin{document}

\title{Analytical Safety Bounds for Trajectory Following Controllers in Autonomous Vehicles
}

\author{\IEEEauthorblockN{Robert Jacumet\rlap{\textsuperscript{1}} , Christian Rathgeber\rlap{\textsuperscript{1}} , Vladislav Nenchev\rlap{\textsuperscript{1}}}
}

\maketitle

\footnotetext[1]{R. Jacumet, C. Rathgeber, and V. Nenchev are with BMW Group, Petuelring 130, 80899 Munich, Germany \texttt{\{robert.jacumet; christian.rathgeber; vladislav.nenchev\}@bmw.de} }

\begin{abstract}
A major challenge in autonomous driving is designing control architectures that guarantee safety in all relevant driving scenarios. Given a safe desired reference trajectory for the vehicle, a trajectory following controller has to ensure that the trajectory is followed with a maximally allowed deviation, even in the presence of external disturbances, such as wind gusts or inclined roads. In this paper, a method for computing upper bounds for linear, time invariant single-input-multiple-output systems with state feedback controllers and additive bounded disturbances is proposed. The bounds for the offset between states and the reference are derived analytically based on worst case disturbance sequences. For systems with two states the bounds are strict, while for higher order systems they are conservative. The method is applied to obtain position bounds for lateral trajectory following controllers, and to analyze how different choices of feedback parameters affect safety margins.
\end{abstract}

\section{Introduction}
\vspace{-11.5cm}
\mbox{\small 
	This~work~has~been~accepted~to~the~2023~9\textsuperscript{th}~International~Conference~on~Control~Decision~and~Information~Technologies~(CoDIT).}
\vspace{10.7cm}

Increasing the automation level of road vehicles brings many benefits such as improved safety, efficiency and comfort.  Despite having attracted a growing interest in research and development \cite{Maurer.2016,Liu.2021}, ensuring safety remains a challenging task for automated driving vehicles, which have to operate in dynamic and uncertain environments, and interact with other vehicles and road users \cite{Eskandarian.2012}.  

A commonly used control architecture for autonomous vehicles is based on a Trajectory Planner (TP) and a Trajectory Following Controller (TFC) \cite{Eskandarian.2012,Rathgeber.2016,Badue.2021}. The TP is responsible for generating a feasible and collision-free motion trajectory for the ego vehicle, taking into account the current state of vehicles, the road geometry, traffic rules, and obstacles \cite{Rathgeber.2016}. The TFC is responsible for compensating external disturbances that may affect the vehicle's motion, such as wind gusts, road slopes or sensor noise such that it follows the trajectory provided by the TP as closely as possible. \cite{Rathgeber.2016}. However, there is no guarantee that the TFC can always follow the trajectory perfectly, due to modeling errors, actuator saturation or \!unexpected large disturbances. Therefore, if the deviation between the actual vehicle state and the desired trajectory exceeds a certain threshold, the vehicle may collide with other objects or leave the road boundaries. This poses a serious threat to the safety of autonomous driving, especially for higher levels of automation where human intervention is minimal or absent~\cite{Rathgeber.2016,Badue.2021}. To address this issue, formal methods can be used to provide rigorous proofs of safety for the TP and TFC based on mathematical models and specifications of the system behavior~\cite{Nenchev2021}. 

In a highly automated driving system, it is necessary to formally guarantee that the deviation to the trajectory does not exceed a specified threshold. Most existing methods for finding bounds on the system's states in the presence of disturbances use set invariance techniques. In \cite{Shuyou.2014,DiCairano.10.12.201313.12.2013},
robust control invariant sets are calculated to capture the safe operation sets of the system. Alternatively, safety verification can be achieved by reachable set estimation \cite{Xiang.2017}. A drawback of set-invariance-based techniques
is that they do not allow to obtain analytical bounds for the maximal offset of the states from the reference dependent on the controller and plant
parameters. Other approaches use the system’s transient response or the worst case disturbance to estimate the deviation, but they do not provide exact bounds \cite{Wang.2022}, \cite{Kogan.1998}. To
the best knowledge of the authors, an analytical bound for the maximal offset under bounded disturbances has not been obtained yet.

In this paper, we propose a method to derive analytical upper bounds on the system states in continuous time, which are exact for systems with up to two states and a conservative over-approximation for higher order systems. An analytical expression of the disturbance sequence causing the largest state values is used to analytically derive upper bounds on the states of single-input-multiple-output (SIMO) systems with feedback
controllers and additive bounded disturbances. The obtained
expressions depend on the controller and plant parameters, allowing to analyze which parameters ensure the requested safety relevant behavior.
The method is applied to find upper bounds on the lateral position and heading angle of a vehicle controlled by a lateral TFC, and to study how different choices of feedback gains affect the safety guarantees.

This paper is organized as follows. Section II introduces the problem statement. The method for finding analytical upper bounds on the system's states in continuous time is derived in Section III and applied to a simplified model of a lateral TFC in Section~IV.  A conclusion is provided in Section V.

\section{Problem Statement}
Consider a bounded disturbance $\vec{z}(t)\!\in\!\mathcal{Z}\!=\![-\vec{z}_\text{max},\vec{z}_\text{max}] \subset \mathds{R}^{n_z}$. To determine analytical bounds on the system's states, we consider SIMO linear time invariant (LTI) system families of the form
\begin{align}
	\dot{\vec{x}}(t) &= \Mat{A}(\vec{q})\vec{x}(t) + \vec{b}(\vec{q}) u(t) + \Mat{E}(\vec{q}) \vec{z}(t),
\end{align}
where $u(t) \in \mathds{R}$ denotes the scalar control signal at time t, $\vec{x}(t) \in \mathds{R}^{n_x}$ are the states, and the input and output matrices are~$\Mat{A}~\in~\mathds{R}^{n_x \times n_x}$, $\vec{b} \in \mathds{R}^{n_x}$, and $\Mat{E} \in \mathds{R}^{n_x \times n_z}$. The system matrices depend on the plant parameters $\vec{q} \in \mathcal{Q} =  [\vec{q}_\text{min},\vec{q}_\text{max}]\subset \mathds{R}^{n_q}$, which are assumed to be fixed over states, inputs, and time.  The control signal is calculated by a linear state feedback controller, i.e., $u(t) = \vec{k}^\top \vec{x}(t)$, where  $\vec{k} \in \mathds{R}^{n_x}$ is chosen such that the resulting closed loop system
\begin{align}
	\dot{\vec{x}}(t) &= \underbrace{(\Mat{A}(\vec{q}) -\vec{b}(\vec{q})\vec{k}^\top )}_{\Mat{A}_\text{CL}(\vec{q})}\vec{x}(t) + \Mat{E}(\vec{q}) \vec{z}(t) \label{eq:frame_CL_system}
\end{align}
is Hurwitz stable. 

Finding the maximal deviation of the $k$-th state in the presence of a bounded disturbance $\vec{z}$ is formulated as the following optimization problem:
\begin{subequations}	\label{eq:frame_WC_optimization}
	\begin{align}
		x_{k,\text{max}}(t) &=  \max_{\vec{z}(t)\in \mathcal{Z}} x_k(t) \label{eq:frame_WC_optimization_cost}\\
		\text{s.t.} \qquad  \dot{\vec{x}}(t) &= \Mat{A}_\text{CL}(\vec{q})\vec{x}(t) + \Mat{E}(\vec{q}) \vec{z}(t) \label{eq:frame_WC_optimization_a}, \\
		\quad \vec{x}(0) &= \vec{0}.\label{eq:frame_WC_optimization_b}	\end{align}
\end{subequations}
As there is no upper limit on time, $\vec{x}(0)=\vec{0}$ is assumed without loss of generality. In the following, the goal is to derive an analytical solution to this optimization problem.
\section{Computing upper bounds analytically}
Using the analytical solution for linear differential equations~\cite{Zeidler.2013},
we can write the optimization problem \eqref{eq:frame_WC_optimization} as
\begin{align}
	x_{k,\text{max}}(t) &=  \max_{\vec{z}(t)\in \mathcal{Z}}  \int_{0}^t \left[ e^{\Mat{A}_\text{CL}(\vec{q})\tau}\Mat{E}(\vec{q}) \vec{z}(t-\tau) \right]_{k} \,d\tau, \label{eq:x_1_general_max_problem}
\end{align}
where $\left[ \vec{a}\right]_{k}$ denotes the $k$-th component of the vector $\vec{a}$. In the following, we do not explicitly write down the matrices' dependencies on $\vec{q}$ anymore. To simplify \eqref{eq:x_1_general_max_problem}, we maximize over each entry of $\vec{z}$ independently:
\begin{align}
	x_{k,\text{max}}(t) &=  \max_{\vec{z}(t)\in \mathcal{Z}}  \int_{0}^t \left[ e^{\Mat{A}_\text{CL}\tau} \sum_{j=1}^{n_z} \vec{e}_jz_j(t-\tau) \right]_{k} d\tau  \nonumber\\
	&= \sum_{j=1}^{n_z}  \max_{z_j(t)\in \mathcal{Z}_j}  \underbrace{ \int_{0}^t  \left[ e^{\Mat{A}_\text{CL}\tau}  \vec{e}_jz_j(t-\tau) \right]_{k}}_{x_{k,j}(t)} \,d\tau,  \label{eq:x_1_general_max_problem_split}
\end{align}
where $\vec{e}_j$ is the $j$-th column of $\Mat{E}$ and $z_j$ is the $j$-th entry of $\vec{z}$. Therefore, the problem of finding the maximizing $\vec{z}$ reduces to finding the vector's entries independently. In the following, we look at one of the summands $x_{k,j}$ in \eqref{eq:x_1_general_max_problem_split}, as the others follow analogously. Let $\vec{z}_\text{WC}(t) = \text{arg}\max_{\vec{z}(t)\in \mathcal{Z}} x_k(t)$ under the constraints \eqref{eq:frame_WC_optimization_a}~-~\eqref{eq:frame_WC_optimization_b} be the worst case disturbance signal causing the maximal state offset. We denote  $\max_{z_j(t)\in \mathcal{Z}_j} x_{k,j}(t)$  with $x_{k,j,\text{max}}(t)$ so that  $x_{k,\text{max}}(t) \! =\! \sum_{j=1}^{n_z} x_{k,j,\text{max}}(t)$.
In order to maximize $x_{k,j}(t)$, the respective integrand found in \eqref{eq:x_1_general_max_problem_split} needs to be maximized by $z_j(t-\tau)$ for $0 \! \leq\! \tau\! \leq \!t$. This is the case if~$z_j(t-\tau) = z_{j,\text{max}} \,\text{sgn}\left(\left[ e^{\Mat{A}_\text{CL}\tau} \vec{e}_j \right]_{k} \right)$ as this leads~to
\begin{align}
	x_{k,j}(t) &\leq z_{j,\text{max}}  \int_{0}^t \left| \left[ e^{\Mat{A}_\text{CL}\tau} \vec{e}_j \right]_{k} \right|\,d\tau = 	x_{k,j,\text{max}}(t). \label{eq:WC_Offset_general}
\end{align}
 Let $\text{sgn}(a) =	+1 \,\,  \text{if} \,\,  a \geq 0$ and $\text{sgn}(a) =	-1 \,\,  \text{if} \,\,  a < 0$.
The disturbance sequence causing the largest $x_{k,j}$ value is switching between~$z_{j,\text{max}}$ and $-z_{j,\text{max}}$, depending on the sign of the integrand, i.e.:
\begin{equation}
	z_{j,\text{WC}}(t-\tau) = z_{j,\text{max}} \,\text{sgn}\left(\left[ e^{\Mat{A}_\text{CL}\tau} \vec{e}_j \right]_{k} \right), \, \forall \, 0 \leq \tau \leq t. \label{eq:frame_WC_sequence_general}
\end{equation}
Due to the absolute value, the integrand \eqref{eq:WC_Offset_general} is always positive and larger than zero. We expect $\vec{e}_j \neq \vec{0}$. As a result, $x_{k,j,\text{max}}(t)$ is strictly increasing with time, though always converging for stable systems. Therefore, we have two types of bounds for the maximal offset. The time dependent bound $x_{k,\text{max}}(t)$ is the worst case offset caused by the additive disturbance over a time period $t$ when starting with zero offset at $t=0$.
The time independent bound $x_{k,\text{max}} = \lim\limits_{t \to \infty} x_{k,\text{max}}(t) $ describes the maximal offset the $k$-th state  can reach under an additive bounded disturbance.
A disturbance acting on the system with its maximal value but no switching, i.e.,~$z_{j,\text{fix}}(t-\tau)  = z_\text{max} \,\, \forall  \,0 \leq \tau \leq t$, leads to 
\begin{equation}
	x_{k,j,\text{fix}}(t) = z_{j,\text{max}}  \int_{0}^t \left[ e^{\Mat{A}_\text{CL}\tau} \vec{e}_j \right]_{k} \,d\tau,
\end{equation}
which, in general, is easier to calculate but only equals $x_{k,j,\text{max}}$  if there are no sign switches of $e^{\Mat{A}_\text{CL}\tau}$ in $\tau \in [0,t]$.

The goal is to solve the integrand in \eqref{eq:WC_Offset_general} analytically.
 The upcoming formulas allow an arbitrary number of real eigenvalues once or twice as a root of the characteristic polynomial, i.e., having an algebraic multiplicity of $1$ or $2$, whereas complex eigenvalues must have an algebraic multiplicity of $1$. Formulas for higher algebraic multiplicities exist as well, but are not common to appear for sufficiently small systems. Since we can enforce the needed multiplicities by the choice of $\vec{k}$, we will not regard them here. Therefore, each of the eigenvalues $\lambda_1,\dots, \lambda_{n_x}$ of $e^{\Mat{A}_\text{CL}\tau}$ falls into one of the following three categories:
\begin{itemize}
		\item Distinct real eigenvalues: $\lambda_i \neq \lambda_j \,\, \forall j\neq i, \lambda_i \in \mathds{R}$;
		\item Double real eigenvalues: $\lambda_i = \lambda_j \in \mathds{R}$ for exactly one~$j\neq i$;
		\item Complex conjugated eigenvalues: $\lambda_i = \lambda_j^*$, for exactly one $j\in \{1,\dots n_x\}$, $ \lambda_i \in \mathds{C}\backslash \mathds{R}$.
	\end{itemize}
The terms $n_\text{r}$, $n_\text{d}$, and $n_\text{c}$ denote the number of distinct real, double real and complex conjugated eigenvalues, respectively. Clearly, $n_x= n_\text{r}+ n_\text{d} + n_\text{c}$ and $n_\text{d}$, as well as $n_\text{c}$ are  even numbers. We sort the eigenvalues such that the first~$n_\text{d}$ are double real eigenvalues, the $n_\text{c}$ complex conjugated eigenvalues follow  pairwise, and the $n_\text{r}$ real and distinct eigenvalues are the last ones.

Due to the previous sorting of the eigenvalues, let:
\begin{equation}
	g_i(\tau) = 
	\begin{cases}
		c_{i} \tau e^{\lambda_{i}\tau} & \text{if} \,\, i \leq n_\text{d},  \,\,i \,\,\text{odd}\\
		c_{i} e^{\lambda_{i-1}\tau} & \text{if}  \,\, i \leq n_\text{d}, \,\, i\,\, \text{even}\\
		c_{i}e^{\lambda_{i}\tau} & \text{if} \,\,  n_\text{d} \leq i \leq n_\text{d}+  n_\text{c}, \,\, i \,\, \text{odd}\\
		c_{i-1}^*e^{\lambda_{i-1}^*\tau} & \text{if} \,\,  n_\text{d} \leq i \leq n_\text{d}+  n_\text{c}, \,\, i \,\, \text{even}\\
		c_{i}e^{\lambda_{i}\tau} & \text{if}  \,\, n_\text{d} +  n_\text{c} \leq i \leq n
	\end{cases}, \label{eq:frame_gi_start}
\end{equation}
where $c_1,\dots, c_n$ are constants.
By performing a Jordan matrix decomposition of the closed loop system matrix, i.e., $	\Mat{A}_\text{CL}~=~ \Mat{P}\Mat{J}\Mat{P}^{-1}$, where $\Mat{J}$ is in Jordan canonical form and $\Mat{P}$ is an appropriate transformation matrix, we find that 
\begin{flalign}
		\left[  e^{\Mat{A}_\text{CL}\tau} \vec{e}_j \right]_{k}
		&=	\left[  \Mat{P}e^{\Mat{J}\tau}\Mat{P}^{-1} \vec{e}_j \right]_{k}= \sum_{i = 1}^{n_x} g_i(\tau). \label{eq:framework_general_eAtb}  
	\end{flalign}
Substituting  \eqref{eq:framework_general_eAtb} into \eqref{eq:WC_Offset_general}  for $x_{k,j,\text{max}}(t)$, gives
\begin{align}
	x_{k,j,\text{max}}(t)  &=   z_{j,\text{max}}   \int_{0}^t \left| \sum_{i = 1}^{n_x} g_i(\tau) \right|\,d\tau. \label{eq:impossible}
\end{align} 
Splitting up the integral at times where the integrand is zero would require to find the potentially infinite zeroes of \eqref{eq:framework_general_eAtb} and is not an option. Thus, in general, \eqref{eq:impossible} cannot be solved analytically.

To find an upper bound for $x_k(t)$ for all times under $z_\text{WC}$, we make use of the subadditivity property $|a+b| \leq |a| + |b|$ and thereby find an upper bound for the maximal offset as
\begin{align}
	x_{k,j,\text{max}}(t) & \leq 	z_{j,\text{max}}  \sum_{i=1}^{n_x/2} \int_{0}^t \left|  g_{2i-1}(\tau) + g_{2i}(\tau) \right| \, d\tau \label{eq:framework_2GB}\\
	& \leq  z_{j,\text{max}}   \sum_{i = 1}^{n_x} \int_{0}^t  \left|  g_i(\tau) \right|\,d\tau.\label{eq:framework_1GB}
\end{align}
The tightness of the upper bound on the maximal offset decreases the more the sum is split by the subadditivity property. While  \eqref{eq:framework_1GB} can be solved analytically, it provides only a conservative bound. As we cannot solve terms of the form $c_1e^{\lambda_1\tau} + 	c_2e^{\lambda_2\tau} + 	c_3e^{\lambda_3\tau} \stackrel{!}{=} 0$ for $\tau$ but can handle $	c_1e^{\lambda_1\tau} + 	c_2e^{\lambda_2\tau}  \stackrel{!}{=} 0$
we tighten the bound by splitting the absolute function over all $g_i$'s into absolute functions over groups of two, as can be seen in  \eqref{eq:framework_2GB}. 
This reduces the problem to solving the three types of integrals that can appear when upper bounding $x_{k,\text{max}}(t)$.

\subsubsection{Real poles case}
In the case of two distinct real poles
$\lambda_i \neq \lambda_j \in \mathds{R}$,  at maximal one single sign switch of the term $c_i e^{\lambda_i\tau} + c_{i+1} e^{\lambda_{i+1}\tau}$ can happen at  $t_s = \frac{1}{\lambda_2 - \lambda_1}\text{log}\left(-\frac{l_1}{l_2}\right)$ if $0<t_s<t$. Acknowledging the potential sign switch, the integral of interest equates to
\begin{align}
	\mu_d (t)\!	&=\!\Big|\frac{c_i}{\lambda_i}(\bar{c}^{\frac{\lambda_i}{\lambda_i-\lambda_j}} \!-e^{\lambda_it}\!-\!1) \!+\! \frac{c_j}{\lambda_j}(\bar{c}^{\frac{\lambda_j}{\lambda_i-\lambda_j}} \!-e^{\lambda_jt}\!-\!1) \Big|\!,\! \label{eq:frame_td_RPs}
\end{align}
with the abbreviation $\bar{c} = \left(-\frac{2c_j}{c_i}\right)$.
The summand for the time independent bound is
\begin{align}
	\mu_d (t\rightarrow \infty)=\Big| \frac{c_i}{\lambda_i}(\bar{c}^{\frac{\lambda_i}{\lambda_i-\lambda_j}}-1) + \frac{c_j}{\lambda_j}(\bar{c}^{\frac{\lambda_j}{\lambda_i-\lambda_j}}-1) \Big|.	 \label{eq:frame_ti_RPs}
\end{align}

\subsubsection{Double real pole case}
In this case, the switch occurs at $t_s = -\frac{c_i}{c_{i+1}}$ if $0<t_s<t$, yielding the summand for the time dependent and time independent bound, respectively. The abbreviation $\nu = c_i - c_{i+1}\lambda_i$ is used.
\begin{align}
	\mu_r (t) &= \frac{1}{\lambda_i^2} \Big|e^{\lambda_i t}( \nu  - c_{i} \lambda_i t ) -2c_{i}e^{-\frac{c_{i+1}}{c_{i}}\lambda_i} + \nu\Big| . \label{eq:frame_td_DRP}\\
\mu_r (t\rightarrow \infty) &= \frac{1}{\lambda_i^2} \Big|-2c_{i}e^{-\frac{c_{i+1}}{c_{i}}\lambda_{i}} + \nu \Big|.  \label{eq:frame_ti_DRP}
\end{align}  

\subsubsection{Complex conjugated poles case}
The eigenvalues $\lambda_i =\sigma + j\omega = \lambda_{i+1}^* \in \mathds{C}$ lead to infinite sign changes of $ c_i e^{\lambda_i\tau} + c_{i}^* e^{\lambda_i^*\tau}$ in $[0,t]$ for $t\rightarrow \infty$. We denote the real part of $\lambda_i$ with $\sigma$ and the imaginary part with $\omega$, where we expect $\sigma <0$ and $\omega>0$. The analytical solution of the integral is presented in Appendix \ref{c:Appendix_Bounds_System2}. Here, only the main result is presented.
With $N(t)$ being the number of zeros of the term $c_i e^{\lambda_i\tau} + c_{i}^* e^{\lambda_i^*\tau}$, $F(t)$ its anti-derivative and $\phi$ being the term's phase:
\begin{align}
	\phi &= \text{arctan}(\frac{\Im\{c_i\}}{\Re\{c_i\}}), \nonumber\\
	N(t) &= \lfloor \frac{1}{\pi}\omega t-\frac{1}{\pi} \Big( \big(\frac{\pi}{2}  -\phi\big)\text{mod}(\pi)\Big) \rfloor, \nonumber\\ 
	F(t) &= \frac{1}{|\lambda_i|^2}e^{\sigma t}	\Big( \sigma\text{cos}(\omega t+\phi) + \omega\text{sin}(\omega t+\phi)\Big),  \nonumber\\
	F(0) &= \frac{1}{|\lambda_1|^2}	\Big( \sigma\text{cos}(\phi) + \omega\text{sin}(\phi)\Big),  \nonumber
\end{align} the time dependent value of the integral is 
\begin{multline}
 \mu_c (t) =    \frac{4|c_i| \omega}{|\lambda_i|^2}     \frac{1-e^{\frac{\sigma}{\omega}\pi(N(t)+1)}}{1-e^{\frac{\sigma}{\omega}\pi}} e^{\frac{\sigma}{\omega}  \Big( \big(\frac{\pi}{2}  -\phi\big)\text{mod}(\pi) \Big) } \\
	+2\text{sgn}(\Re\{c_i\})|c_i|\Big(  (-1)^{N(t)} F(t) - F(0) \Big), \label{eq:frame_td_CCP}
\end{multline}
and the time independent value of the summand is
\begin{align}
	&  \frac{4|c_i| \omega}{|\lambda_i|^2}   \frac{1}{1\!-\!e^{\frac{\sigma}{\omega}\pi}}  e^{\frac{\sigma}{\omega}  \Big(\! \big(\!\frac{\pi}{2}  -\phi\!\big)\text{mod}(\pi)\!\Big)} \!-\! 2\text{sgn}(\Re\{c_i\}\!)|c_i| F(0). \label{eq:frame_ti_CCP}
\end{align}
After solving the three types of integrals potentially appearing in the upper bound of $x_{k}$, we get the time dependent bound by using the formulas \eqref{eq:frame_td_RPs}, \eqref{eq:frame_td_DRP}, and \eqref{eq:frame_td_CCP} in~\eqref{eq:framework_2GB} to get all $x_{k,j,\text{max}}$ appearing in $x_{k,\text{max}}(t) \! =\! \sum_{j=1}^{n_z} x_{k,j,\text{max}}(t)$.  For the time independent version, we use \eqref{eq:frame_ti_RPs}, \eqref{eq:frame_ti_DRP}, and \eqref{eq:frame_ti_CCP}.

In the following, we obtain the upper bound for a trajectory following controller, assuming a double integrator system.

\section{Application to a trajectory following controller}
\label{c:Application_to_TFC_Model}
To demonstrate the developed method for calculating bounds on the system's states, it is applied to a simplified model of a lateral trajectory following controller. The state vector $\vec{x} = \begin{bmatrix}
	\Delta d & \Delta \theta
\end{bmatrix}^\top$   contains the vehicle's lateral offset $\Delta d$ and track angle error $\Delta \theta$.  The control input $u$ is a curvature and $w$ is the reference curvature signal provided by the trajectory planner. The vector of controller parameters is $\vec{k}= \begin{bmatrix}	K_d & K_\theta \end{bmatrix}^\top$, and the vector of plant parameters, here containing only the velocity, is $\vec{q} =  \begin{bmatrix}	v  \end{bmatrix} \in  \mathcal{Q} = \left[   v_\text{min} ,  v_\text{max}\right]$, where $v_\text{min} >0$ and $ v_\text{max}<\infty$. Last, $z$ is an additive disturbance on the curvature and is bounded to $\mathcal{Z}=[-z_\text{max},+z_\text{max}]$, representing e.g., crosswinds. The double integrator system
\begin{align}
	\dot{\vec{x}}  = \begin{bmatrix} 0 & v \\ 0 & 0 \end{bmatrix}  \vec{x}  +  \begin{bmatrix} 0  \\ v \end{bmatrix}  u -  \begin{bmatrix} 0  \\ v \end{bmatrix}  w +  \begin{bmatrix} 0  \\ v \end{bmatrix}  z \label{eq:SS_2n_states}
\end{align}
describes the  vehicles lateral movement relative to a reference, given by the trajectory planner. Using the feedback controller  $u = w - \begin{bmatrix}	K_d & K_\theta \end{bmatrix} \vec{x}$ yields the state equations of the closed loop model in state space representation:
\begin{align}
	\vec{\dot{x}} &=\!  \begin{bmatrix}0 & v \\-vK_d& -vK_\theta \end{bmatrix} \vec{x}  +    \!      \begin{bmatrix} 0  \\v   \end{bmatrix}z 
	= \Mat{A}_{\text{CL}}(\vec{q}) \vec{x} + \vec{e}(\vec{q}) z. \label{eq:2n_SS_CL_states}
\end{align}
 
Utilizing the method proposed in Sec III, to find analytical upper bounds on the position offset $\Delta d$ from the required reference, we solve
\begin{align}
	x_{1,\text{max}}(t) &=  \max_{z(t)\in \mathcal{Z}}  \int_{0}^t \left[ e^{\Mat{A}_\text{CL}(\vec{q})\tau}\vec{e}(\vec{q}) z(t-\tau) \right]_{1} \,d\tau \label{eq:example}\\
	&=  z_\text{max}  \int_{0}^t \left| \left[  e^{\Mat{A}_\text{CL}\tau} \vec{e} \right]_{1} \right|\,d\tau. \label{eq:2n_x1_general_formula_integral}
\end{align}
Using the inverse Laplace transform 
$ e^{\Mat{A}_\text{CL}\tau} \vec{e}  \!=\!  \mathcal{L}^{-1}\{(s\Mat{I}\!-\!\Mat{A}_\text{CL})^{-1}\}\vec{e}$ or the Jordan decomposition shown in \eqref{eq:framework_general_eAtb}, we~get
\begin{align}
	\left[  e^{\Mat{A}_\text{CL}\tau} \vec{e} \right]_{1} = 
	\begin{cases} 
	 v^2 \tau  e^{\lambda_1 \tau}    &\text{if}  \,\, \lambda_1 = \lambda_2\\
		\frac{v^2}{\lambda_1 - \lambda_2 } e^{\lambda_1 \tau} + \frac{v^2}{\lambda_2 - \lambda_1 } e^{\lambda_2 \tau}  & \text{if} \,\, \lambda_1 \neq \lambda_2 \\
	\end{cases}. \label{eq:2n_x1_general_formula_integrand}
\end{align}
To solve \eqref{eq:2n_x1_general_formula_integral}, it has to be determined whether the system has two double real eigenvalues, two distinct real eigenvalues or a pair of complex conjugated ones. If $\lambda_1\in \mathds{R}, \lambda_2 \in \mathds{R}$ and $\lambda_1 \neq \lambda_2$ is chosen, then
\begin{align}
	g_1(\tau) &= \frac{v^2}{\lambda_1 - \lambda_2 } e^{\lambda_1 \tau} \qquad \text{ and }\qquad
	g_2(\tau) = \frac{v^2}{\lambda_2 - \lambda_1 } e^{\lambda_2 \tau}. \nonumber
\end{align}
The time independent bound on the positional offset is obtained by substituting the above values into   \eqref{eq:frame_ti_RPs}, yielding
\begin{align}
	x_{1,\text{max}}  =   z_\text{max}  \int_{0}^t \left| g_1(\tau) + g_2(\tau) \right|\,d\tau = z_\text{max}\frac{v^2}{\lambda_1 \lambda_2}.    \nonumber
\end{align}
If $\lambda_1\in \mathds{R}, \lambda_2 \in \mathds{R}$ but $\lambda_1 = \lambda_2$ is chosen, then
\begin{align}
	g_1(\tau) &= v^2\tau e^{\lambda_1 \tau} \qquad \text{ and }\qquad
	g_2(\tau) =0. \nonumber
\end{align}
Using \eqref{eq:frame_ti_DRP},  the time independent bound for a double real eigenvalue is
\begin{align}
	x_{1,\text{max}}    = z_\text{max} \int_{0}^\infty  \left| g_1(\tau)\right| + \left| g_2(\tau)\right|\,d\tau = z_\text{max}\frac{v^2}{\lambda_1^2} \nonumber.
\end{align}
If $\lambda_1 = \sigma + j\omega = \lambda_2^* \in \mathds{C} \backslash \mathds{R}$, we have a pair of complex conjugated eigenvalues. This leads to
\begin{align}
	g_1(\tau) &= -\frac{v^2j}{2\omega} e^{\lambda_1 \tau} \qquad \text{ and }\qquad
	g_2(\tau) = \frac{v^2j}{2\omega} e^{\lambda_1^* \tau} \nonumber
\end{align}
and, using \eqref{eq:frame_ti_CCP}, the time independent bound is
\begin{align}
	x_{1,\text{max}} =   z_\text{max}  \int_{0}^t \left| g_1(\tau) + g_2(\tau) \right|\,d\tau =  -z_\text{max}\frac{v^2}{|\lambda_1|^2} \frac{e^{\frac{\sigma}{\omega}\pi} + 1}{e^{\frac{\sigma}{\omega}\pi} - 1}. \nonumber
\end{align}
Note that $x_{1,\text{max}}$ is a tight bound for the system offset. In case of the worst case disturbance sequence acting on the system, $x_1(t)$ will reach $x_{1,\text{max}} $ for $t \rightarrow \infty$.

The eigenvalues can be expressed as functions of the controller parameters, from which  it can be seen that $K_\theta^2 \geq 4K_d$ leads to distinct real eigenvalues, $K_\theta^2 < 4K_d$ to complex conjugated eigenvalues and an equality to double real eigenvalues:
 \begin{align}
	\lambda_{1,2} = -\frac{v}{2}\left(K_\theta \pm \sqrt{K_\theta^2-4K_d}\right).  \label{eq:2n_EValues}
\end{align}
By using \eqref{eq:2n_EValues}, the exact maximal possible position offsets can be written as functions of the controller and plant parameters:
\begin{align}
	x_{1,\text{max}}  \!=\! &
	\begin{cases}
		\!\frac{z_\text{max}}{K_d} &  \text{if} \,\, K_\theta^2 \geq 4K_d\\
		\!\frac{z_\text{max}}{K_d}  \!\left(    \frac{2}{1-e^{-\frac{ K_\theta}{\sqrt{4K_d - K_\theta^2}}\pi}} -1 \!\right)\!& \text{if} \,\,  K_\theta^2 < 4K_d \\
	\end{cases}. \label{eq:2n_bound2_k_end}
\end{align}

\begin{figure*}[ht]
	\centering
	\includegraphics[width=0.78\textwidth]{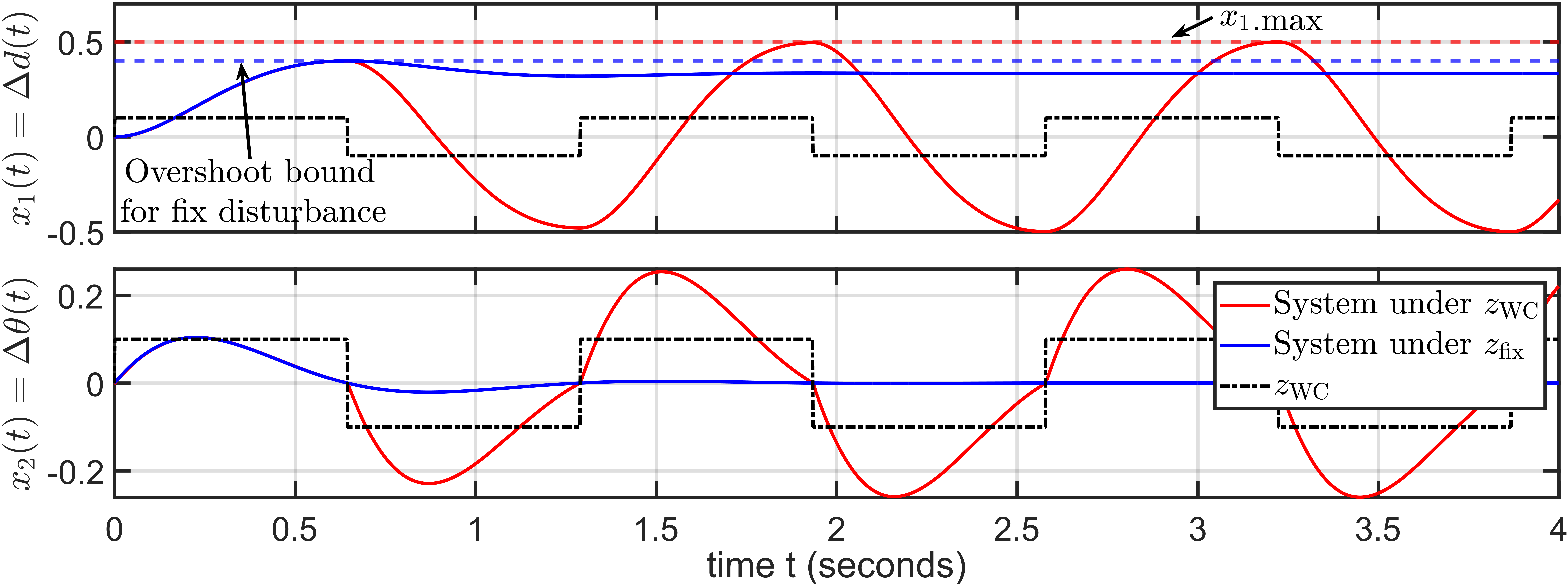}
	\caption[State trajectory of the $n_x=2$ system under a fixed disturbance and the worst case disturbance.]{ The state trajectories under a fixed (blue) and worst case  disturbance sequence (red) are shown. The values $(z_\text{max},K_d,K_\theta,v) = (0.1,0.3, 0.5,10)$ are used. The  dashed red line is the analytical bound $x_{1,\text{max}}$. The blue dashed line from literature \cite{ Keviczky.2019}  only bounds the overshoot for a constant disturbance.}
	\label{fig:2n_complex_bound_overshoot}	
\end{figure*}

Figure \ref{fig:2n_complex_bound_overshoot} shows the system's states under the worst disturbance sequence and under a fixed disturbance $z_\text{fix}= z_\text{max}$, for the case of two complex eigenvalues, i.e., when~$K_\theta^2 < 4K_d$ is chosen. Here, $v=10 \,\frac{\text{m}}{\text{s}}$ and $z_\text{max}=0.1 \,\frac{\text{1}}{\text{m}}$ is selected. The dashed blue line denotes the maximal position offset for $z_\text{fix}$, which is an overshoot before returning to the equilibrium $\frac{z_\text{max}}{K_d}$. The formula for a maximal overshoot in double integrating systems for constant inputs can be found in \cite{Keviczky.2019,DAzzo.2003}  and is
\begin{equation}
	\frac{z_\text{max}}{K_d}\left( e^{\frac{-\xi\pi}{\sqrt{1-\xi^2}}}+1\right),
\end{equation}
with $\xi = -\frac{\Re\{\lambda_1\}}{|\lambda_1|}$in our case. For evaluating the worst case offset under a bounded disturbance, the expressions derived in Sec. III can be used.

\begin{figure}[ht]
	\centering
	\includegraphics[width=0.467\textwidth]{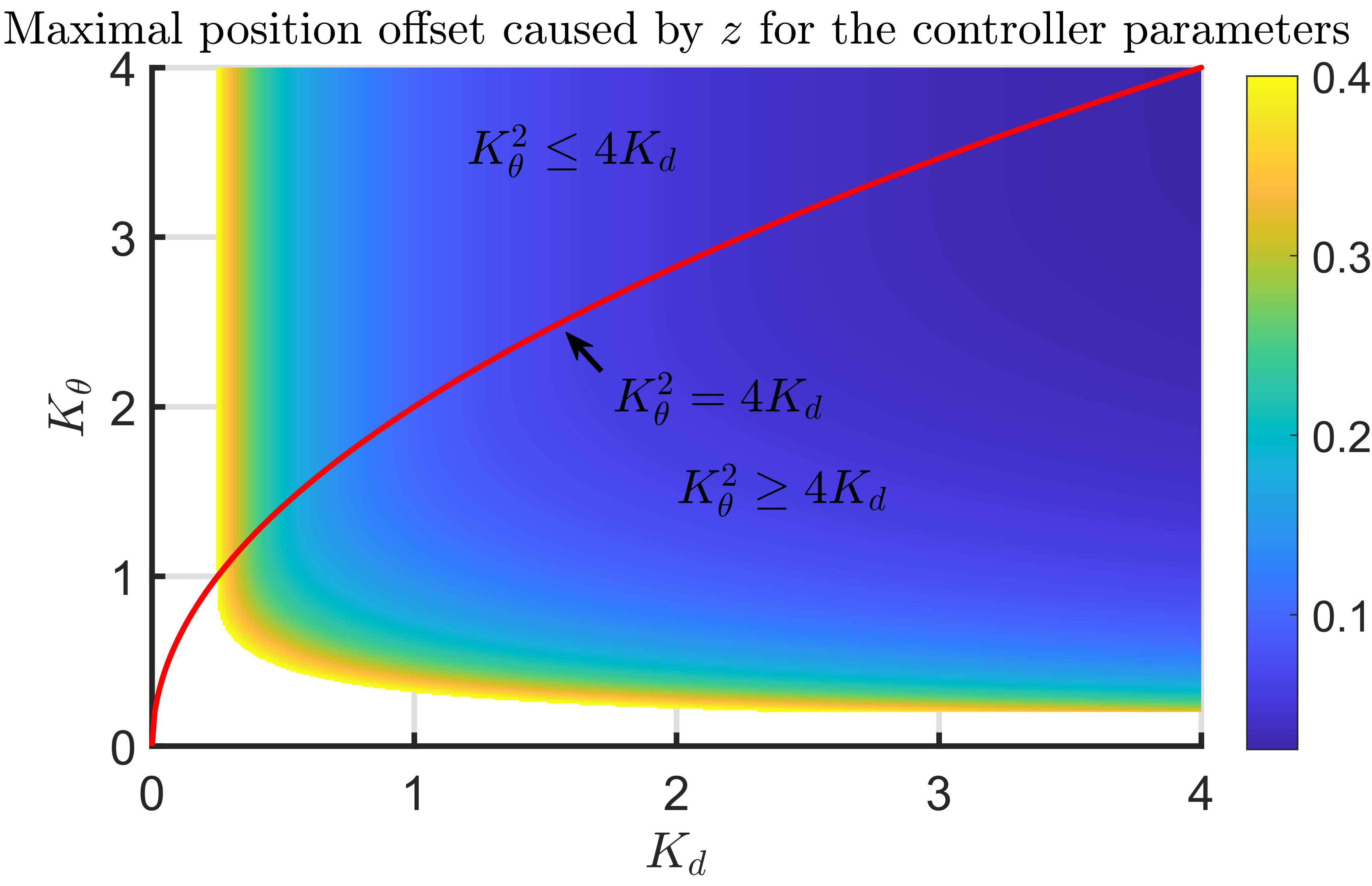}
	\caption[Controller parameter dependent maximal position offsets for the $n_x=2$ system.]{ The maximal position offsets are color-coded and shown dependent on the choice of controller parameters for $(z_\text{max},v) = (0.1,10)$. Only $K_d$ and $K_\theta$ values  that lead to $x_1(t)  \leq  \Delta d_\text{max}= 0.4 \,$m are colored.  Additionally, it is marked where complex conjugated or real eigenvalues are present.  }
	\label{fig:2n_max_k}	
\end{figure}

Finally, we use \eqref{eq:2n_bound2_k_end} to calculate the controller parameters which guarantee the position offset to be ~$x_1(t)  \leq  \Delta d_\text{max}$ for all times $t$. Figure \ref{fig:2n_max_k} shows the resulting $x_{1,\text{max}}$ as a function of the controller parameters~$K_d$ and $K_\theta$, with $v=10 \,\frac{\text{m}}{\text{s}}$ and $z_\text{max}=0.1 \,\frac{\text{1}}{\text{m}}$. For $\Delta d_\text{max}$ we use the value $0.4 \,\text{m}$.

The analysis shows how the derived analytical bounds can be used to pick parameters which guarantee that a maximal offset in position is not exceeded, hence fulfilling the safety requirement of the lateral TFC.
\section{Conclusion}

In this work, we proposed a method to compute analytical upper bounds for states of linear SIMO systems with bounded, additive disturbances in continuous time. To this end, the disturbance sequence causing the largest state values was derived and used to solve an integral analytically. Depending on the type of the system's eigenvalues, we consider three cases. Analytical expressions have been derived for each case. For systems with a state dimension of $2$, the proposed approach provides tight bounds on the maximal state values. These analytical bounds differ from peak overshoot bounds of second order systems, as they give the maximal state values for bounded arbitrary input signals and not only for fixed signals acting on the system. 

The developed method was applied to a simplified model of a lateral TFC to find which control parameters guarantee that the position offset between the vehicle and a desired reference is smaller than a pre-defined value.

Due to the general formulation, the application of the approach is not limited to TFC architectures only. Future work may deal with finding guarantees on the tightness of the proposed offset bounds for systems with states of dimensions greater 2.

\appendices

\chapter{Solving the Complex Conjugated Poles Case}
\label{c:Appendix_Bounds_System2}
The main challenge in solving
\begin{equation}
	\mu_c(t) = \int_{0}^t  \left| c_i e^{\lambda_i\tau} + c_{i}^* e^{\lambda_i^*\tau} \right| d\tau \label{eq:appendix_start_integral}
\end{equation} 
with $\lambda_i = \lambda_{i+1}^* = \sigma + j\omega$, $\sigma <0$, $\omega>0$, and $c_i = c_{i+1}^* \in \mathds{C}\backslash \mathds{R}$ is the infinite sign changes of $ c_i e^{\lambda_i\tau} + c_{i}^* e^{\lambda_i^*\tau}$ in $[0,t]$ for~$t$ approaching infinity, caused by the imaginary exponential functions being a decaying oscillation. Therefore, the integral~\eqref{eq:appendix_start_integral} needs to be split up into possibly infinitely many sections at the integrand's zeroes, in order to deal with the absolute function.

First, the complex term is rewritten as a linear combination of a cosine and a sine, having only real coefficients, multiplied by a real, decaying exponential function:	
\begin{align}
	c_i e^{\lambda_i\tau} + \!c_{i}^* e^{\lambda_i^*\tau} &= 	2e^{\sigma \tau} \Big(\Re(c_i) \text{cos}(\omega \tau)- \Im(c_i)\text{sin}(\omega \tau)\Big) \nonumber\\
	&= 2e^{\sigma \tau} \Big(\text{sgn}(\Re\{c_i\})|c_i|\text{cos}\Big(\omega \tau + \phi\Big)\!\Big)  \label{eq:appendix_one_cos},
\end{align}
which is possible for sinusoidal linear combinations with equal frequency and phase \cite{Zeidler.2013}. The abbreviation $\phi = \text{arctan}(\frac{\Im\{c_i\}}{\Re\{c_i\}})$ is used, so that including the absolute value in \eqref{eq:appendix_one_cos} gives
\begin{align}
	\left| c_i e^{\lambda_i\tau} + c_{i}^* e^{\lambda_i^*\tau} \right| &= 2|c_i|
	e^{\sigma \tau}  \Big|\text{cos}\Big(\omega \tau + \phi \Big)\Big|. \label{eq:appendix_one_cos_abs}
\end{align}
Integrating over \eqref{eq:appendix_one_cos_abs} cannot be done directly, but for the integrand  \eqref{eq:appendix_one_cos} without the absolute function, we make use of the closed analytical formula for its antiderivative \cite{Zeidler.2013}:
\begin{align}
	&\int_{t_0}^{t}	e^{au+c}\text{cos}(bu+\phi) du \nonumber\\
	&= \Big[\frac{1}{a^2 + b^2}e^{au+c}	\Big( a\text{cos}(bu+\phi) + b\text{sin}(bu+\phi)\Big)\Big]_{t_0}^t. 
\end{align}
Hence, the antiderivative of $	f(\tau)= 	e^{\sigma \tau}  \text{cos}\Big(\omega \tau  + \phi\Big) $
is determined to be:
\begin{align}
	F(\tau)&=\frac{1}{|\lambda_i|^2}e^{\sigma \tau}	\Big( \sigma\text{cos}(\omega \tau+\phi) + \omega\text{sin}(\omega \tau+\phi)\Big).
\end{align}
Using these terms, the solution of the integral in the absence of the absolute value operation becomes
\begin{align}
	\int_{0}^t \! c_i e^{\lambda_i\tau} \!+ c_{i}^* e^{\lambda_i^*\tau} d\tau 
	&=  2|c_i|\text{sgn}(\Re\{c_i\}) \int_{0}^{t} f(\tau)d\tau \nonumber\\ 
	&=  2|c_i|\text{sgn}(\Re\{c_i\}) \Big(\!F(t)\!-\! F(0)\!\Big)\!,
\end{align}
where we made use of the Newton-Leibniz axiom \cite{Zeidler.2013}.
The goal now is to solve the integral \eqref{eq:appendix_start_integral}, i.e.
\begin{align}
	\int_{0}^t  \left| c_i e^{\lambda_i\tau} + c_{i}^* e^{\lambda_i^*\tau} \right| d\tau 
	&=  2|c_i| \int_{0}^{t} \left|f(\tau) \right| d\tau. \label{eq:appendix_integral_to_solve_intermediate}
\end{align}
Assume there is $N+1$ zeros of $f(\tau)$ in $[0,t]$ at $0\leq n_0 <n_1 <n_2 <\dots<n_N\leq t$. Their locations are determined by
\begin{equation} 
	n_k = \frac{1}{\omega}  \Big( \big(\frac{\pi}{2}  -\phi\big)\text{mod}(\pi) +k\pi \Big), \quad k = 0,1,\dots,N \label{eq:appendixero_locations}
\end{equation}
and mark the values where the integral \eqref{eq:appendix_integral_to_solve_intermediate} is split, as $f$ changes its sign.
The areas in which $f$ is negative need to be taken into account with a minus due to the absolute function. We first assume that $f$ is negative in $[0,n_0]$ and $N$ is odd for a fixed $t$. This way we can rewrite $\int_{0}^{t} \left|f(\tau) \right| d\tau$ as
\begin{align}
	 &-\int_{0}^{n_0}	f(\tau) d\tau - \sum_{i=1}^{N}\int_{n_{i-1}}^{n_{i}} (-1)^{i}	f(\tau) d\tau 
	- \int_{n_N}^{t}	f(\tau) d\tau \nonumber\\ 
	 & =  F(0) -F(t) +2 \sum_{i=0}^{N} (-1)^{i+1} F(n_i). \label{eq:appendix_integral_split_general}
\end{align}

By comparing \eqref{eq:appendix_integral_split_general} with $\int_{0}^{t} f(\tau)d\tau = F(t)-F(0)$, two things can be observed. Without the absolute function and the sign switches introduced by it, only the antiderivatives evaluated at $0$ and $t$ appear in the integral's solution. With the absolute function on the other hand, antiderivatives evaluated at the zeroes, and therefore possibly infinitely many terms, appear with an alternating sign. Second, in~\eqref{eq:appendix_integral_split_general}, the signs in front of  $ F(0)$ and~$ F(t)$ depend on the signs of $f(\tau)$ in $[0,n_0)$ and $(n_N,t]$, respectively. More specifically, the sign in front of $ F(0)$ follows $-\text{sgn}(f(0)) = -\text{sgn}(\Re\{c_i\})$. When $N(t)+1$ is even, $F(t)$ goes into \eqref{eq:appendix_integral_split_general} with the same sign as $F(0)$. The number of zeroes $N(t)+1$, which we from now on see as time dependent, can be calculated using
\begin{equation}
	N(t) = \lfloor \frac{1}{\pi}\omega t-\frac{1}{\pi} \Big( \big(\frac{\pi}{2}  -\phi\big)\text{mod}(\pi)\Big) \rfloor.
\end{equation}
Therefore, in the general case, where we do not need to assume the sign of $f$ in $[0,n_0]$ or that the number of zeroes is odd or even, the term $F(0)-F(t) $ in \eqref{eq:appendix_integral_split_general} can be replaced by 
\begin{align}
	\text{sgn}(\Re\{c_i\})\big((-1)^{N(t)}F(t)  - F(0)\big). \label{eq:appendix_integral_split_general_VZ_corners}
\end{align}

Noting that the antiderivatives' values at the zeroes are
\begin{align}
	F(n_{\text{pos} \rightarrow \text{neg}}) &= -\frac{\omega}{|\lambda_i|^2} e^{\sigma n_{\text{pos} \rightarrow \text{neg}}}, \label{eq:appendix_complex_antiderivative_pos_negero} \\
	F(n_{\text{neg} \rightarrow \text{pos}}) &= \frac{\omega}{|\lambda_i|^2} e^{\sigma n_{\text{neg} \rightarrow \text{pos}}}  \label{eq:appendix_complex_antiderivative_neg_posero},
\end{align}
where $n_{\text{pos} \rightarrow \text{neg}}$ is a zero where the cosine switches from positive to negative function values and $n_{\text{neg} \rightarrow \text{pos}}$ marks a zero of opposite sign switch. It can be seen that the form of \eqref{eq:appendix_complex_antiderivative_pos_negero}  and \eqref{eq:appendix_complex_antiderivative_neg_posero} only varies in a switch of signs. As the antiderivatives at the  $n_{\text{pos} \rightarrow \text{neg}}$ locations also appear in \eqref{eq:appendix_integral_split_general} with a negative sign, the alternating signs in \eqref{eq:appendix_integral_split_general} cancel out. By using \eqref{eq:appendix_integral_split_general_VZ_corners}, \eqref{eq:appendix_complex_antiderivative_pos_negero}, and \eqref{eq:appendix_complex_antiderivative_neg_posero} in \eqref{eq:appendix_integral_split_general}, the solution of the integral simplifies to
\begin{align}
	\int_{0}^{t} \!\left|f(\tau) \right| d\tau \!=&  \text{sgn}(c_i)\big(\!(-1)^{N(t)}F(t)  \!-\! F(0)\!\big) \!+\! \frac{2\omega}{|\lambda_1|^2} \sum_{k=0}^{N} \! e^{\sigma n_k}\!, \label{eq:appendix_integral_sum}
\end{align}
while simultaneously not requiring assumptions on zeroes and the signs of $f$ anymore, in order for the antiderivative at the boundaries $0$ and $t$ to appear with correct signs. By using \eqref{eq:appendixero_locations} in the last term of \eqref{eq:appendix_integral_sum} and rewriting the result as
\begin{align}
	\frac{2\omega}{|\lambda_1|^2} \sum_{k=0}^{N}  e^{\sigma n_k}   \nonumber &= \frac{2\omega}{|\lambda_1|^2}  e^{\frac{\sigma}{\omega}  \Big( \big(\frac{\pi}{2}  -\phi\big)\text{mod}(\pi) \Big) }   \sum_{k=0}^{N}  e^{(\frac{\sigma}{\omega}\pi)k},
\end{align}
the sum is a geometric series. Using that $ 0<e^{\frac{\sigma}{\omega}\pi}<1$, due to~$\sigma$ being negative in stable systems and $\omega >0$ by definition, we can make use of the formulas 
\begin{equation}
	\sum_{k=0}^{N} ar^k = a\Big(\frac{1-r^{N+1}}{1-r}\Big), 	\qquad 	\sum_{k=0}^{\infty} ar^k = a\Big(\frac{1}{1-r}\Big),
\end{equation}
for the geometric series \cite{Zeidler.2013}. Finally, the solution of \eqref{eq:appendix_start_integral} is
\begin{multline}
	  \frac{4|c_i|\omega}{|\lambda_i|^2}   \frac{1-e^{\frac{\sigma}{\omega}\pi(N(t)+1)}}{1-e^{\frac{\sigma}{\omega}\pi}} e^{\frac{\sigma}{\omega}  \Big( \big(\frac{\pi}{2}  -\phi\big)\text{mod}(\pi) \Big) }   \\
	+2\text{sgn}(\Re\{c_i\})|c_i|\Big(  (-1)^{N(t)} F(t) - F(0) \Big), \nonumber
\end{multline}
and the time independent version is
\begin{align}
	&  \! \frac{4|c_i|\omega}{|\lambda_i|^2}    \frac{1}{1\!-\!e^{\frac{\sigma}{\omega}\pi}}  e^{\frac{\sigma}{\omega}  \Big( \big(\!\frac{\pi}{2}  -\phi\big)\text{mod}(\pi)\!\Big)} - \!2\text{sgn}(\Re\{c_i\})|c_i| F(0)\nonumber.
\end{align}

\bibliographystyle{ieeetr}
\bibliography{refs/mybib}

\ifdefined\AddMyGloss
\glsaddall
	\AddMyGloss 
\fi
\end{document}